\documentstyle[12pt]{article}
\input math_macros.tex

\def\ref#1{$^{#1)}$}
\begin{document}
\begin{titlepage}
\begin{center}
     \hfill    JSUHEP-930701, 1993\\

\vskip .05in

{\large \bf Topological Study of Global Quantization in Non-
Abelian Gauge Theories}
\vskip .05in
Huazhong Zhang\footnote{E-mail: ZHANGHZ@SSCVX1.SSC.GOV,
CPZHANG@FNALV.BITNET}\\[.1in]
{\em Department of Physics and Atmospheric Sciences\\
     P.O.Box 17660\\
     Jackson State University\\
     Jackson, MS 39217}
\end{center}

\begin{abstract}
We study some topological aspects of non-abelian gauge
theories intimately connected to the Lie algebras of the
gauge groups and the homotopy theory in the generalized
gauge orbit space. The physics connection to the
non-perturbative solution to strong CP problem as originally
proposed by the author is also discussed. Some relevant
topological formulas are also given and discussed. A result
from the physics application is that the usual gauge orbit
space on the compactified space can contain at most a $Z_2$
monopole structure in the SP(2N) gauge theories. Some
relevance to the open universe is also discussed. We expect
that our results may also be useful to the other studies of
non-abelian gauge theories in general.
\end{abstract}
\end{titlepage}
\newpage
\renewcommand{\thepage}{\arabic{page}}
\setcounter{page}{1}
\section{Introduction}
The topological study of field theories has been of
fundamental interest since the discovery of Yang-Mills
theories$^1$. In fact, one of the most interesting features
of field theories is the topological and non-perturbative
aspect in non-abelian gauge theories such as instantons$^{2-
3}$ and magnetic monopoles$^{4-6}$. By considering the
possible non-vanishing curvature flux in gauge orbit spaces
and its relevant global quantization, a non-perturbative
solution to the strong CP problem has been proposed
originally$^7$ by the author. The main purpose of this paper
is to study some topological issues intimately connected to
this new proposed solution. Some further physics
applications will also be given. We expect that our study
and the results will be also useful to the other studies of
gauge theories, especially to the general study of the
topological aspects in non-abelian gauge theories.

It is well known that, in non-abelian gauge theories a
$\theta$ term,
\begin{equation}
{\cal L}_{\theta}=\frac{\theta}{32{\pi}^2}
\epsilon^{\mu\nu\lambda\sigma}
F^a_{\mu\nu}F^a_{\lambda\sigma},
\end{equation}
can be added to the lagrangian density of the system due to
instanton effects. The $\theta$ term with an arbitrary value
of $\theta$ can induce CP violation. Especially, such an
effective $\theta$ term in QCD can induce CP violations in
strong interactions. Since we will only focus to the
topological study, the perturbative effects are not so
relevant to our major discussions, the $\theta$ in this
paper may always be regarded physically as denoting the
effective vacuum angle. For convenience in discussions
relevant to the QCD, the $\theta$ should be notationally
regarded as the usual $\theta+arg(detM)$ with M being the
quark mass matrix when the effects of electroweak
interactions are included. It has been an experimental fact
that the strong CP violation can be only very small or
vanishing. This requires that the values of $\theta$ can be
only very special, for example very small $\theta\leq 10^{-
10}$ or vanishing$^8$. However, there is no obvious reason
why the values of $\theta$ have to be so constraint. This is
the well known strong CP problem in elementary particle
physics. As it is well known that one of the most
interesting approaches to solve the strong CP problem is the
assumption of an additional Peccei-Quinn $U(1)_{PQ}$
symmetry$^{10}$. The spontaneous breaking of the $U(1)_{PQ}$
symmetry implies the existence of scalar particles called
axions$^{10-11}$. The essential idea in this solution is
that the existence of axions and their interactions
with the other particles will ensure the vanishing of the
vacuum angle $\theta$ dynamically. However, the experimental
observations have been given more and more constraints on
the possible models for axions$^8$. Although the axion
approach has been plausible and interesting, the fact that
the experiments have not given any support$^8$ to the axions
is an indication that we may need other mechanism to provide
solution to the strong CP problem. Our proposed solution$^7$
is essentially due to the topological application to the
global understanding of quantum theory.

In this paper, we will especially study some topological
results and some physics applications intimately connected
to our proposed solution$^7$. We will also give a discussion
relevant to the open universe in this approach. This paper
will be organized as follows. In next section, we will first
give a brief description of the main results in the proposed
solution$^{7}$ by using the Dirac quantization condition in
the relevant gauge orbit spaces for the vacuum angle
$\theta$ in the presence of magnetic monopoles. With this
preparation, in section 3, we will then discuss and derive
some relevant topological formulas as well as some physics
applications. In section 4, we will discuss the relevance to
the open universe in our approach. Our conclusions will be
given in section 5.

\section{Topological quantization in the relevant gauge
orbit space for the vacuum angle $\theta$}

It is known$^{12-13}$ that gauge theories can be studied in
the Schroedinger formulation with constraints of Gauss' law.
It can be shown that (See Ref.12, for example) the
Hamiltonian equation in the functional space in this
formulation gives the same partition functional in the usual
definition. In this formulation, it is known that the wave
functionals are cross sections of a line bundle on the
relevant gauge orbit space. Namely, the existence of a
well-defined wave functional physically corresponds to the
existence of a cross section in the relevant fiber bundle
topologically. The existence of such a cross section in the
line bundle requires that if the flux of the field strength
or curvature two form through any closed surfaces in the
relevant gauge orbit space is non-vanishing, then it must be
quantized according to Dirac quantization rule.

Our original approach consists of two main generalizations.
The first one is to generalize the concept of gauge orbit
space on a compactified space to that on an open space with
the existence of magnetic monopoles in order to include the
boundary effects consistently. Usually, the physical
configuration space is the quotient space of all the well-
defined gauge potentials modulo the gauge transformations
with continuous gauge functions approaching to the identity
element of the gauge group at the spatial infinity. Namely,
it is the gauge orbit space on the compactified space. We
call this as the usual gauge orbit space$^7$. In the
presence of magnetic monopoles with non-vanishing magnetic
charges, we must include the space boundary in our physics
consideration. The space is topologically a large three-
dimensional ball with the boundary which is topologically a
two sphere $S^2$. We call the quotient space of all the
well-defined gauge potentials restricted on the space
boundary $S^2$ modulo all the well-defined gauge
transformations which are only implemented on the space
boundary as the restricted gauge orbit space$^7$. As it is
known that$^{29}$ the well-defined gauge transformations on
the entire space boundary are those commutive with the
magnetic charges. The physical configuration space is
expected to be the union of the usual and the restricted
gauge orbit spaces. Physically, one needs to understand this
as following. The physical wave functional in the finite
space region is required invariant only under the gauge
transformations approaching to the identity of the gauge
group, or it is a cross section in the relevant line bundle
restricted to the usual gauge orbit space. The physical wave
functional on the space boundary is required invariant only
under the gauge transformations well-defined and implemented
only on the space boundary. Namely, it is a cross section in
the line bundle restricted to the restricted gauge orbit
space. The whole physical wave functional is the product of
both.

Our second generalization has been mainly the extension of
the method in Ref.13 in Schroedinger formulation for the
discussion of an abelian structure inside the non-abelian
gauge theories with a Pontryagin or $\theta$ term. The
formalism with different methods have also been used to
derive the mass parameter quantization in three-dimensional
Yang-Mills theory with a Chern-Simons term$^{12,13}$. Our
approach has been more connected to the method in Ref.13 due
to the explicit abelian structure with non-trivial
topological properties in the gauge orbit spaces constructed
in the Ref.13 is more relevant to our discussion.

Especially, we extended the discussions in Ref.13 for non-
abelian theories to the case in the presence of magnetic
monopoles. We showed that due to the existence of an abelian
field induced by the $\theta$ term in the presence of
magnetic monopoles, there exits non-vanishing flux
proportional to $\theta$ for the corresponding induced field
strength in the restricted gauge orbit space. To have a
well-defined physical wave functional, Dirac quantization
condition ensures that the vacuum angle $\theta$ must be
quantized. We will only give a brief description of the main
results here for the consistency of the paper. One of the
main equations we obtained$^{7}$ is given by
\begin{equation}
\int_{S^2}\hat{\cal F}
=\frac{\theta}{2\pi^2}tr\int_{S^2}\{f\int_{C^1}\delta gg^{-1}\}.
\end{equation}
Where the two sphere $S^2$ on the left side of the equation
is in the relevant gauge orbit space and the $S^2$ on the
right side denotes the space boundary. The f is the monopole
field strength 2-form, and $\hat{\cal F}$ is the projection
of the relevant field strength in the gauge configuration
space to the restricted gauge orbit space given by
\begin{equation}
\hat{\cal F}=\frac{\theta}{4\pi^2}\int_{S^2}tr(\delta a\delta a),
\end{equation}
as an integration over the space boundary $S^2$ with
$\delta a$ being the relevant parameter differentiation of
the gauge potential $a$ in the relevant gauge orbit space.

We derived Eq.(2) through explicit calculation.
Topologically, the above equation can be understood as
follows. By using the relevant exact homotopy sequence$^7$,
it is shown that the equation corresponds to the topological
result given by
\begin{equation}
\Pi_2({\cal U}/{\cal G})\cong\Pi_1({\cal G}).
\end{equation}
Where ${\cal U}/{\cal G}$ is a restricted gauge orbit space
with ${\cal U}$ being the space of induced gauge potentials
A on the space boundary $S^2$, and ${\cal G}$ being the
space of gauge functions restricted to the well-defined
subgroup of the gauge group G and on the space boundary
$S^2$. Namely, the ${\cal G}$ is the space of continuous
gauge transformations which are on the space boundary only.
It is well known that these gauge transformations are
commutive with the magnetic charges$^{29}$.

For the evaluation of the above homotopy group for the
quantization of the $\theta$, we will use the following
topological theorem with a proof given in section 3.

Theorem. Let ${\cal G}$ be the space consisting of all the
continuous mappings g from a $S^2$ to a Lie group G, then we have
\begin{equation}
\Pi_1({\cal G})\cong\Pi_1(G)\oplus\Pi_3(G).
\end{equation}
As it is shown that$^{7}$ for our purpose, the relevant
well-defined gauge subgroup commutive with magnetic charges
can be regarded as an abelian group. To derive the
quantization rule for non-vanishing $\theta$, we only need
to consider the U(1) group as given in ref. 7. Since
$\Pi_3(U(1))=0$ and  $\Pi_1(U(1))=Z$, and to obtain the
relevant topological number in our discussion corresponding
for the U(1) subgroup, we only need to consider the
parameter differentiation $\delta g$ for the mappings g
which depends only on the relevant parameter for the loop
$C^1$. Note that we use the same notation G both for the
gauge group and the well-defined gauge subgroup on the space
boundary, the meaning is clear in the corresponding
discussions. Physically, $\Pi_1(G)$ is non-trivial for the
exact gauge group G outside the magnetic monopole is the
necessary condition for the existence of magnetic monopoles
in a spontaneously broken gauge theory$^{5-6}$. In the
realistic case with $G=SU(3)\times U(1)$, this condition is
satisfied, when the $SU(3)\times U(1)$ is regarded as
the exact gauge symmetry of a spontaneously broken gauge
theory, for example SU(5), as it is known that the minimal
SU(5) model has monopole solutions$^{31}$.

Now for the quantization of $\theta$ implied, note that the
left side of Eq.(2) is the magnetic flux of the monopole in
the relevant restricted gauge orbit space. The case of
vanishing $\theta$ always ensures the vanishing flux.
We are interested in case with the possible non-vanishing
vacuum angle $\theta$ in the theory. Using Dirac
quantization condition with magnetic monopoles$^{4-6}$, we
obtained our quantization rule$^{7}$ for the $\theta$ given
by
\begin{equation}
\theta=0, \frac{2\pi}{n}~~(n\neq 0).
\end{equation}
Where the integer n is the effective topological charge of
the generalized monopole$^{15}$ which can be written
as$^{7}$
\begin{equation}
n=-2<\delta',\beta>=\sum_{i=1}^{r}n_i
=-2\sum_{i=1}^{r}\frac{<\alpha_i,\beta>}{<\alpha_i,\alpha_i>},
\end{equation}
with the magnetic charge up to a conjugate transformation by
a group element written in the form of
\begin{equation}
\int_{S^2}f=G_0=4\pi\sum_{i=1}^{r}\beta^{i}H_{i},
\end{equation}
Where $\{H_i\mid i=1, 2,...,r=rank(G)\}$ form a basis for
the Cartan subalgebra$^{30}$ of the gauge group G with
simple roots $\alpha_i$ (i=1,2,...,r). The minus sign is due
to our normalization convention for Lie algebra generators
$tr(L_aL_b)=-\frac{1}{2}\delta_{ab}$.
In general, if the well-defined gauge subgroup on the space
boundary consists of multiple U(1) factors giving multiple
non-vanishing $n_i$, the $\theta$ will be vanishing.
As we noted that$^7$, up to a conjugate transformation, we
can always choose that all the $n_i$'s have the same sign if
non-vanishing. Physically, we may expect that in the single
monopole case, only the fundamental monopoles with $n=\pm 1$
are stable.

Therefore$^{7}$, the existence of magnetic monopoles with
$n=\pm 1$, or $n\geq 2\pi 10^9$ as well as others to ensure
the vanishing of $\theta$ can provide solution to the strong
CP problem due to the global quantization of the effective
$\theta_{QCD}$ . This concludes the brief description of our
main results for the quantization of $\theta$.

\section{Relevant topological theorems}
In this section, we will give a proof to the topological
theorem we stated in the section 2. The corresponding
generalized theorems will also be given, since it may be
useful for the other more general studies of topological
applications in physics. As another physics application, we
will study the possible monopole structure in the usual
gauge orbit space.

Let ${\cal G}=G^{S^2}=\{f\mid f:S^2\rightarrow G ~continuous\}$ be
the space consisting of all the continuous mappings f from a
two sphere $S^2$ to a Lie group G, then we will show that
the fundamental group of ${\cal G}$ is given by $\Pi_1({\cal
G})\cong\Pi_1(G)\oplus\Pi_3(G)$.

Proof: It is well-known that we have the following
fibration$^{14}$
\begin{equation}
P:~{\cal G}\rightarrow G.
\end{equation}
Where the projection P is defined by $P(f)=f(x_0)$ with
$x_0\in S^2$ is the base point. The fiber of this fibration
is given by
\begin{equation}
P^{-1}=\{f\mid f\in{\cal G}~ and~ f(x_0)=e\}
=\{f\mid f:(S^2, x_0)\rightarrow (G,e)\},
\end{equation}
where e denotes the identity element of the Lie group G.
Thus, by definition the fiber $P^{-1}$ is the second order
loop space of G, i.e.
\begin{equation}
P^{-1}(e)=\Omega^2G.
\end{equation}
Where for simplicity, the base point is not written
explicitly since the relevant homotopy groups in our
discussions for the loop spaces based on different base
points are isomorphic for the Lie groups$^{14}$.
The relevant exact homotopy sequence for this fibration is
of the form
\begin{equation}
\Pi_N(G)\stackrel{\Delta_*}{\longrightarrow}
\Pi_{N-1}(\Omega G)
\stackrel{i_*}{\longrightarrow}\Pi_{N-1}({\cal G})
\stackrel{P_*}{\longrightarrow}\Pi_{N-1}(G)~~(N\geq 1).
\end{equation}
The case of N=2 corresponds to our discussions for the
fundamental group of ${\cal G}$.
One can easily see that the fibration
$P:{\cal G}\rightarrow G$ has a cross section$^{14}$
$s:G\rightarrow{\cal G}$ defined by
\begin{equation}
s(g)(x)=g ~for~ any~ x\in S^2, g\in G.
\end{equation}
Where a cross section is a continuous mapping
$s:G\rightarrow{\cal G}$ such that Ps(g)=g for any $g\in G$.
According to the well-known splitting theorem$^{14}$, the
existence of a cross-section of a fiber bundle implies the
splitting of the corresponding exact homotopy sequence.
Namely, the homotopy group of the bundle space can be
written as the direct sum of the corresponding homotopy
groups for the base space and the fiber. Therefore, we have
\begin{equation}
\Pi_N({\cal G})\cong\Pi_N(\Omega^2G)\oplus\Pi_N(G)~~(N\geq 1).
\end{equation}
Especially in the case of N=1, we obtain
\begin{equation}
\Pi_1({\cal G})\cong\Pi_1(G)\oplus\Pi_1(\Omega^2G).
\end{equation}
Now according to the well-known isomorphism relation$^{14}$,
we have
\begin{equation}
\Pi_1(\Omega^2G)=\Pi_2(\Omega G)=\Pi_3(G).
\end{equation}
Therefore, the theorem is proved. The corresponding more
general theorem can be proved similarly and is given by

Theorem 1. Let ${\cal G}$ be the space consisting of all the
continuous mappings from a $S^n~(n\geq 1)$ to a Lie group G, then we have
\begin{equation}
\Pi_N({\cal G})\cong\Pi_N(G)\oplus\Pi_{N+n}(G)~~(N\geq 1).
\end{equation}

There is another relevant theorem$^{14}$ which is also
useful to our discussions. We will include it here and give
a relevant physics application.

Theorem 2. Let ${\cal G}$ be the space consisting of all the
continuous mappings from a $S^n~(n\geq 1)$ to a Lie group G,
such that a given point in the $S^n$ is always mapped to the
identity element of G. Then we have
\begin{equation}
\Pi_N({\cal G})\cong\Pi_{N+n}(G)~~(N\geq 1).
\end{equation}

This theorem is essentially due to the isomorphism
relation$^{14}$
\begin{equation}
\Pi_N(\Omega^{M+1}G)=\Pi_{N+1}(\Omega^MG).
\end{equation}

As a physics application of the theorem 2, we will consider
the possible monopole structures in the usual gauge orbit
space. Denote by ${\cal G}$ all the gauge transformations
mapping the spatial infinity to the identity element of the
gauge group, namely, they are mappings from the compactified
space $S^3$ to the gauge group. The possible monopole
structure is determined by the homotopy group given by
\begin{equation}
\Pi_1({\cal G})\cong\Pi_{4}(G),
\end{equation}
due to the theorem 2. For simple and compact gauge group G,
it is known that the $\Pi_4(G)=Z_2$ is non-vanishing only
for SP(2N) ($SP(2)\cong SU(2)$). Therefore, there can be at
most a $Z_2$ monopole structure in the usual gauge orbit
spaces for SP(2N) gauge theories. Note that since the
relevant homotopy group is finite, it cannot be realized as
by the finite flux of the curvature in the gauge orbit space
and therefore, it cannot give a finite constraint on the
vacuum angle in the usual gauge orbit space. This is
consistent with the fact that in the usual gauge theories on
the compactified space without axion, the $\theta$ angle is
arbitrary.

As a remark in this section, note that the exact homotopy
sequences have also been used to the study of global (non-
perturbative) gauge anomalies$^{16-24}$ and
supersymmetry$^{25}$, especially in term of the James
numbers of Stiefel manifolds.

\section{Relevance to the Open Universe}
In this section, we will give a brief discussion for the
relevance of our results to the topology of the universe. In
our discussions, the boundary effects of the magnetic
monopoles are explicitly involved, if monopole solution is
the true solution to the strong CP problem, then the
boundary effects on the topological properties of the
universe needs to be considered. In fact, our discussions
and results have important consequences on the structure of
the universe.

As we have seen in our discussions and results in the
previous sections that for quantizing the vacuum angle
$\theta$, a non-vanishing magnetic flux through the space
boundary $S^2$ is needed in the presence of magnetic
monopoles. This implies that the universe must be open if
the monopoles provide the true solution to the strong CP
problem, otherwise, the relevant magnetic flux can be only
vanishing.

Our discussions in the previous sections are consistent with
this physical implication. In the realization of the
relevant topological numbers corresponding to $\Pi_1(G)$ for
the well-defined gauge subgroup G, as we have seen that$^7$
only gauge functions depending on the parameter for the loop
$C^1$ in Eq.(2) are needed. Essentially for the relevant
well-defined gauge subgroup this is also the topological
number corresponding to the homotopy group $\Pi_1({\cal G})$
for the relevant space of ${\cal G}$ for the well-defined
gauge transformations on the space boundary. This does not
mean that the space boundary can be identified as single
point. In fact, one can easily see that our result implies
that if non-contractable loops $C^1$ exist with $n\neq 0$,
then the Dirac quantization condition ensures that the space
boundary cannot be continuously contracted into a single
point.

This can be seen as follows. Since as we have seen that in
the section 2, the projection $\hat{\cal F}$ in the relevant
gauge orbit space is given in term of an integration on the
space boundary $S^2$, the quantization of the non-vanishing
flux on the left side of Eq.(2) by the Dirac quantization
condition ensures that the space boundary cannot be
contracted continuously into a single point, otherwise the
quantized non-vanishing flux corresponding to the
non-trivial topological number for the relevant
$\Pi_2({\cal U}/{\cal G})$ can be continuously reduced to
zero flux corresponding to the trivial topological number
for the $\Pi_2({\cal U}/{\cal G})$ which is a contradiction
to the homotopy inequivalence of the different topological numbers.
Therefore, if magnetic monopoles provide the
solutions to the strong CP problem, the space boundary
cannot be continuously contracted into a single point, there
exists non-vanishing magnetic flux through the space
boundary and consequently the universe must be open. The
open universe in the presence of non-vanishing magnetic
charges can be regarded as a quantum effect of the global
quantization.

\section{Conclusions}
In this paper, we have studied some topological issues
intimately connected to the Lie algebras and homotopy
theory. We discussed about the applications to the
topological aspects of non-abelian gauge theories with a
$\theta$ term. Especially, we have discussed about the non-
perturbative solution to the strong CP problem with magnetic
monopoles as originally proposed by the author. The vacuum
angle can be ensured vanishing by non-vanishing topological
charges corresponding to the multiple U(1) factor in the
well-defined gauge subgroup on the space boundary. The
strong CP problem can also be solved with non-vanishing
$\theta$, due to the existence of magnetic monopoles of
topological charge $n=\pm 1$ or $\mid n\mid\geq 2\pi 10^9$.
However, with non-vanishing $\theta$, the CP in strong
interactions cannot be exactly conserved since under CP in
the relevant cases of our discussions, $n\rightarrow -n$ and
the different topological charges n correspond to different
monopole sectors or different physical systems. In fact, as
we conjuctured$^7$ that the CP violation in weak
interactions may be intimately connected to the effects of
magnetic monopoles. Since the CP symmetry is violated in the
realistic world, we may expect that the strong CP is not
exactly conserved. From this consideration, if the strong CP
problem is solved due to the magnetic monopoles, then we
expect that it is actually solved by the quantization of the
non-vanishing vacuum angle.

We note that the $\theta$ term with the existence of
magnetic monopoles was first considered$^{26}$ relevant to
$U_A(1)$ problem and chiral symmetry. It is noted by Witten
that$^{27}$ t' Hooft and Polyakov monopoles with an
arbitrary $\theta$ will carry electric charges proportional
to $\theta$. This was also generalized$^{28}$ to the case of
generalized magnetic monopoles$^{15}$. As we have
emphasized$^{7}$, only non-singular magnetic monopole can
provide solution to the strong CP problem. For example, it
is known that minimal SU(5) model has smooth monopole
solutions$^{31}$.

Moreover, we have also derived and discussed about some
relevant topological formulas. A result from the physics
application is that the usual gauge orbit space on the
compactified space can contain at most a $Z_2$ monopole
structure in SP(2N) gauge theories. We expect that the
topological formulas may also be useful to the other studies
of non-abelian gauge theories in general.

We have also discussed about the fact that the universe must
be open if the monopoles provide the solution to the strong
CP problem. This may be regarded as a quantum effect of the
global quantization. Therefore, the fact that the strong CP
violations can be only so small may imply the existence of
magnetic monopoles and the universe must be open. The
understanding of the strong CP problem may provide
information for spacetime structure and the structure of our
universe.

Acknowledgement: The author would like to express his
gratitude to Y. Li, Profs. K. Bardakci, S. Okubo, Y. S. Wu,
A. Zee and other people in the Theoretical Physical Group at
Lawrence Berkeley Laboratory for valuable discussions.

\end{document}